\documentstyle[twoside, epsf]{article}

\input ibvsx.sty

\begin{document}
\IBVShead{5923}{1 February 2010}

\IBVStitle{The Polar CSS~081231:071126+440405 at a Low Accretion Rate}

\IBVSauth{THORNE$^1$, K., GARNAVICH$^1$, P., MOHRIG$^1$, K. }

\IBVSinst{University of Notre Dame, Notre Dame, IN 46556, e-mail: kthorne@nd.edu}

\SIMBADobjAlias{CSS~081231:071126+440405}
\IBVStyp{CV/polar}
\IBVSkey{photometry}
\IBVSabs{Time-resolved photometry of the polar CSS~081231:071126+44040 was obtained with the VATT}
\IBVSabs{in October 2009. The light curve shows a single accretion hotspot lagging 10 degree behind}
\IBVSabs{the secondary. The last two nights of data show ``sputtering'' accretion and
the hotspot nearly disappears in consecutive cycles.}

\begintext

{\bf Abstract:} Time-resolved photometry of the eclipsing polar CSS~081231:071126+44040
was obtained with the Vatican Advanced Technology Telescope (VATT) on four nights
in October 2009. The light curve shows a single accretion hotspot lagging 10$^\circ$
behind the secondary. The last two nights of data show ``sputtering'' accretion
and the hotspot nearly disappears in consecutive orbits.

\vspace{0.5cm}

CSS~081231:071126+440405 (AAVSO Alert Notice \#142) is a suspected
polar that went into bright outburst
in early 2009. A polar is an accreting white dwarf with a strong magnetic field
that disrupts the accretion disk, funneling material directly on
to the magnetic poles. CSS~081231:071126+440405 shows deep
eclipses with a period of about 1.94 hours and reached a peak brightness
of $V\sim 14.8$ in March, 2009 (AAVSO Special Notice \#149). 

We imaged CSS~081231:071126+440405 with
the Vatican Advanced Technology Telescope (VATT) over four consecutive
nights 2009 Octocber 22-25 (UT) using the VATT4K CCD. The CCD was
binned by two pixels and only the first 512 pixels were read out,
reducing the overhead to 10~sec. We continuously took 30~sec exposures in the $V$ band
spanning 3.5 hours on the first two nights, then switched to 20~sec
exposures in the $B$-band for the final two nights. Using images of
the Landolt (1992) standard region SA113, we estimated
a zeropoint for the VATT photometry. For the star at the USNO-B.1 coordinates
$\alpha =$7:11:22.839, $\delta =$+44:04:12.45 (30 arcsec west of the
variable) we find $V=16.57\pm0.05$ mag and $B-V$=0.36 mag.

\IBVSfig{9.0cm}{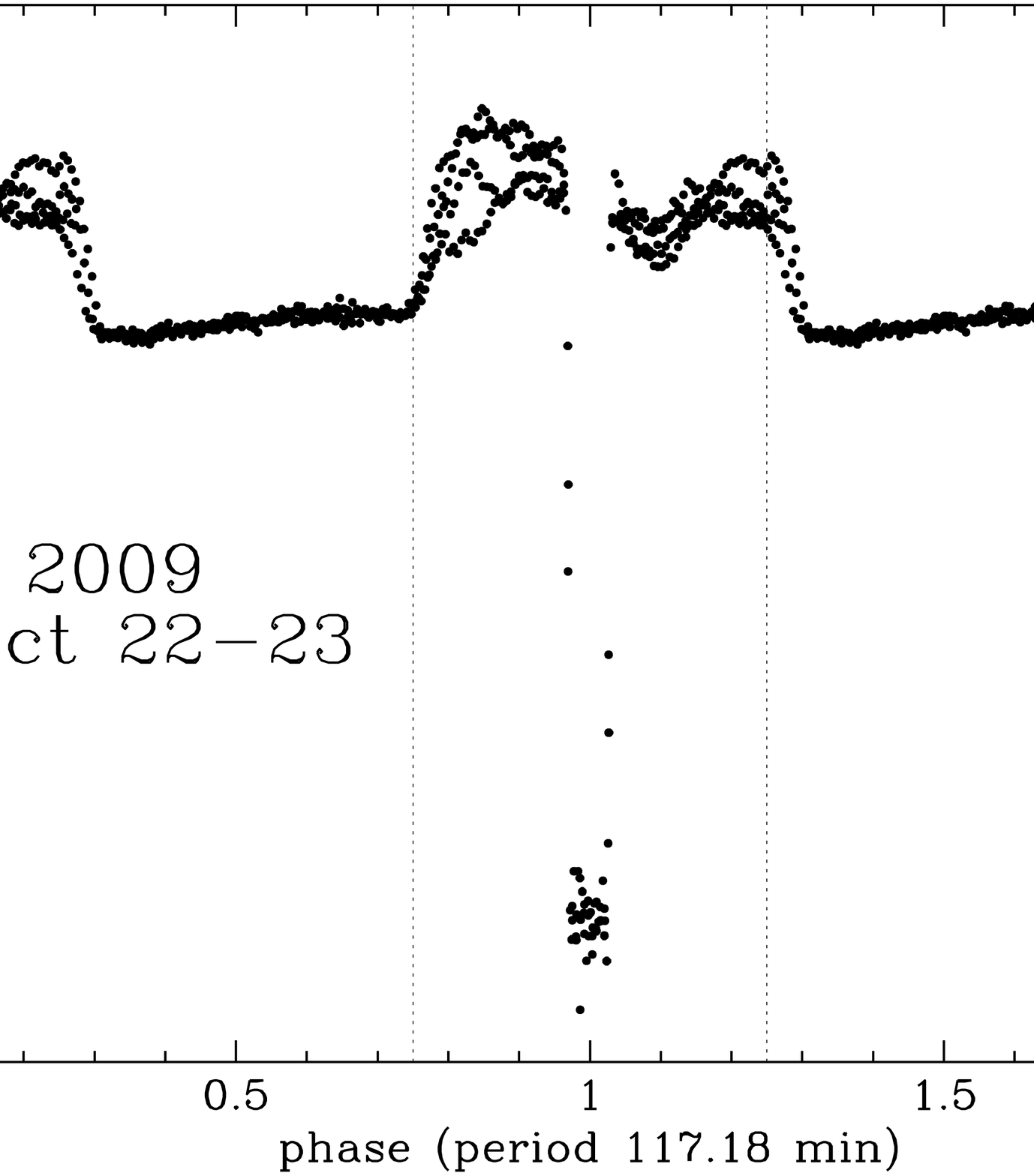}{
The phased $V$-band light curve from the first two nights of VATT observations.
The accreting hotspot is visible on each side of the eclipse but is occulted
by the white dwarf for half the orbit. The dotted lines mark phases $\pm90^\circ$ 
to show that the hotspot lags the companion by 10$^\circ$.}

The $V$-band light curve is shown Figure~1. Two short, deep eclipses are
clearly seen each night and the heliocentric times of mid-eclipse
are given in Table~1. We derived a period of 117.181$\pm 0.004$ minutes and
used it to phase the photometric data. The eclipse times in Table~1 provide
an ephemeris in heliocentric Julian days of  
$${\rm HJD}=2455126.8960(1)+{\rm E}\times 0.081376(3) $$ where
the numbers in parentheses are the uncertainties on the final decimal place.
We find the full eclipse length is 0.058$\pm 0.001$ in phase or 6.80$\pm 0.12$ minutes.

The light curve from Oct. 22-23 shows a bright plateau between phases
$-0.25$ and $+0.25$ surrounding the eclipse. During
the plateau, the star is strongly variable but shows a consistent
dip in brightness near phase +0.1
which is likely to be self-absorption by the accretion column.
Between phases $+0.25$ and $+0.75$ the star displays a slow, steady rise
of 0.1~mag and a brightness that is extremely consistent from orbit to orbit.

The light curve is similar to the eclipsing polar HU~Aqr in its low accretion
state (Schwope et al. 2001). We expect CSS~081231:071126+440405 has 
a single hotspot on the accreting white dwarf which is in synchronous rotation
with the secondary star. The hotspot
is occulted by the white dwarf for half the spin period, opposite the
phase of the eclipse, suggesting that the accreting magnetic pole
is nearly facing the secondary star.
A careful look at the light curve shows that the occultation of
the hotspot is shifted by 10.5 degrees (0.03 in phase) relative to
the eclipse. This implies that the hotspot trails the line between
the primary and secondary stars by about 10$^\circ$. In HU~Aqr, the accretion spot
leads the secondary by 30$^\circ$ to 50$^\circ$.

The time it takes for the plateau phase to rise to full brightness or
disappear depends on the size of the hotspot as it is revealed or
blocked by the white dwarf limb. The
hotspot latitude and vertical displacement also affect the timing of
the hotspot occultation (Schwope et al. 2003). We estimate the
ingress/egress of the hotspot takes about 0.05$\pm 0.01$ in phase.
While the hotspot is occulted there is a 10\% rise in brightness
suggesting that temperature varies with longitude on the white dwarf.
The color of the system while the hotspot is occulted is $B-V=0.17\pm0.02$ mag.

Figure~2 shows that the $B$-band phased light curve from Oct. 24-25
differs significantly from the previous two nights. While the
plateau from the hotspot is present during the first orbit each night,
it is essentially gone on the second cycle. This suggests the mass
transfer is ``sputtering'' as it ends an active accretion phase. The
star is three magnitudes fainter than its peak in 2009 March, and
the accretion may be becoming sporadic at this low rate. 

Assuming the eclipse is total, we estimated the brightness and color of the
secondary star. At minimum the star is very faint, so the
eight to ten individual short exposures during each eclipse were added
together to improve the signal-to-noise ratio. The secondary star's brightness
is $V=20.86\pm0.05$ and $B=22.6\pm0.2$, consistent with a
late M-type dwarf star (note that reddening in this direction is as much as
E($B-V$)=0.075 mag (Schlegel et al. 1998)). Correcting for the contribution
of the secondary star, the color of the white dwarf
plus accretion stream is $B-V$=0.09 mag, but at this low accretion rate the
light is likley dominated by the white dwarf.


\IBVSfig{9.0cm}{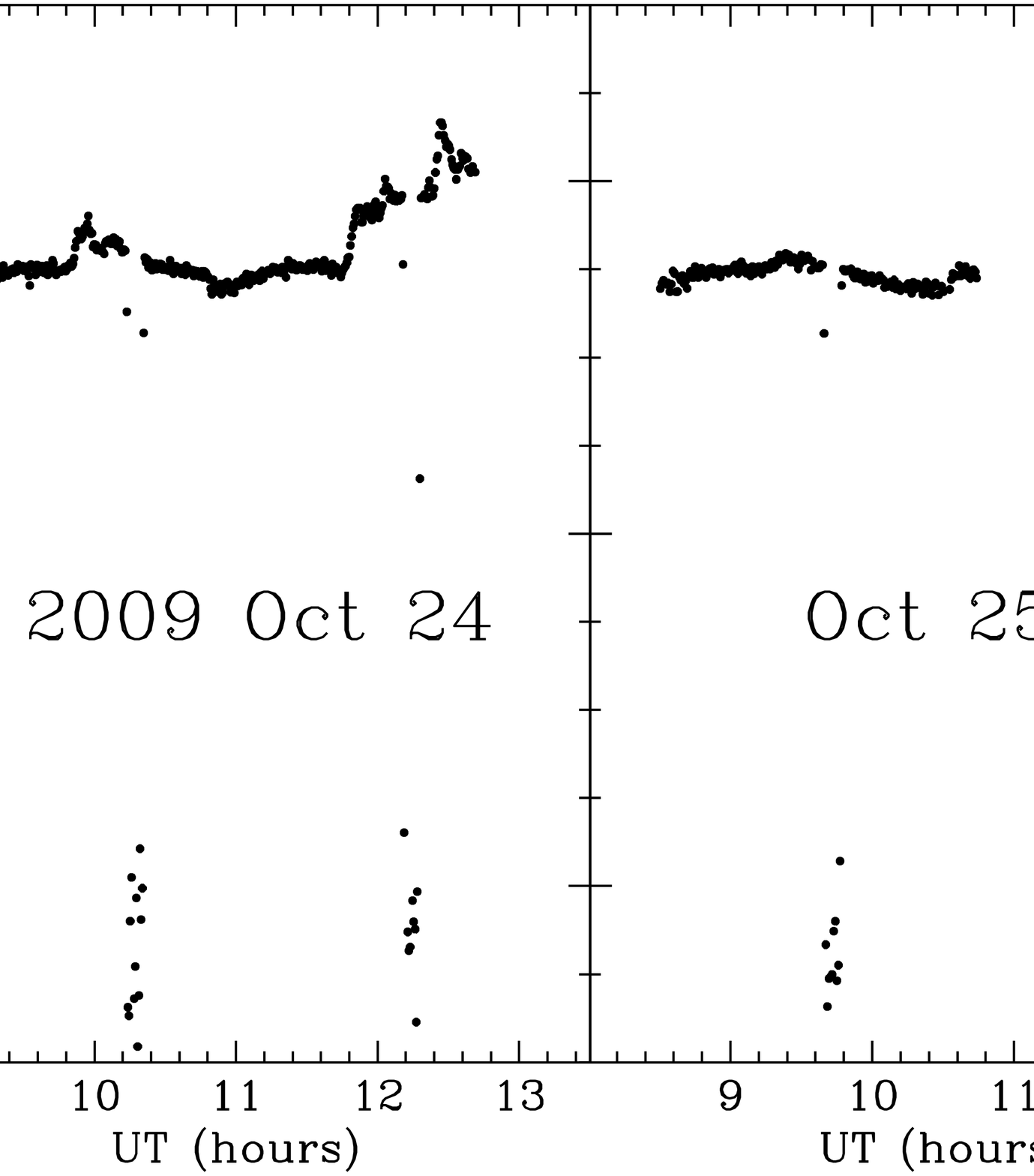}{
The $B$-band light curves from the last two nights of VATT observations. A gap in the
Oct. 25 light curve around 11~UT was caused by clouds. On both nights
the hotspot is very weak during the first orbit but has recovered  on the
second orbit suggesting
the mass transfer is becoming sporadic.}

\begin{center}
{Table 1. Observed Times of Mid-Eclipse}
\vskip 1mm
\begin{tabular}{ccccc}
\hline
Date   & Bandpass &  Epoch  &  HJD & error \\
(UT)   &          &         &      & (days) \\
\hline
2009 Oct. 22 & $V$ & 0 & 2455126.8960  & 0.0001 \\
2009 Oct. 22 & $V$ & 1 & 2455126.9773  & 0.0001 \\
2009 Oct. 23 & $V$ & 12 & 2455127.8724  & 0.0001 \\
2009 Oct. 23 & $V$ & 13 & 2455127.9539  & 0.0001 \\
2009 Oct. 24 & $B$ & 25 & 2455128.9303  & 0.0001 \\
2009 Oct. 24 & $B$ & 26 & 2455129.0117  & 0.0001 \\
2009 Oct. 25 & $B$ & 37 & 2455129.9069  & 0.0001 \\
2009 Oct. 25 & $B$ & 38 & 2455129.9883  & 0.0002 \\
\hline
\end{tabular}
\end{center}

\vskip 0.5cm

\references

Landolt, A., 1992, {\it AJ}, {\bf 104}, 340

Schlegel, D., et al. 1998, {\it ApJ}, {\bf 500}, 525


Schwope, A. D., Schwarz, R., Sirk, M. and Howell, S. B. 2001, {\it A\&A}, {\bf 375}, 419

Schwope, A. D., et al. 2003, {\it A\&A}, {\bf 402}, 201

\endreferences

\end{document}